\documentclass[prb,twocolumn,aps,showpacs]{revtex4}
\usepackage{graphicx}
\usepackage{bm}
\usepackage{amssymb} 
\begin{document}

\title{Disorder-induced tail states in a gapped bilayer graphene}

\author{V. V. Mkhitaryan and M. E. Raikh}

\affiliation{ Department of Physics, University of Utah, Salt Lake
City, UT 84112}

\begin{abstract}

The instanton approach to the in-gap fluctuation states
is applied  to the spectrum of biased bilayer graphene.
It is shown that the density of
states falls off  with energy measured from the band-edge as
$\nu(\epsilon)\propto
\exp(-\vert\epsilon/\epsilon_t\vert^{3/2})$, where the
characteristic tail energy, $\epsilon_t$, scales with the
concentration of impurities, $n_i$, as $n_i^{2/3}$. While
the bare energy spectrum is characterized by two energies:
the bias-induced gap, $V$, and interlayer tunneling, $t_{\perp}$,
the tail, $\epsilon_t$, contains a {\it single} combination
$V^{1/3}t_{\perp}^{2/3}$. We show that the above expression
for   $\nu(\epsilon)$    in the tail actually applies all the way down to the
mid-gap.

\end{abstract}
 \maketitle

\section{Introduction}
Several experimental studies of electronic properties of graphene
bilayers were recently reported in the literature \cite{EOhta06,
ENovoselov06, ESavchenko07, EGap07, Ehop07, EMobility08,
EStormer08, EPinczhuk08}. While experiments Refs.
\onlinecite{ENovoselov06, ESavchenko07, EMobility08, EStormer08,
EPinczhuk08} were carried out on unbiased, and thus
gapless\cite{Falko06} bilayers, the focus of the papers Refs.
\onlinecite{EOhta06, EGap07,Ehop07} was the fact that a tunable
gap emerges in the energy spectrum of bilayer upon applying an
interlayer bias \cite{Falko06,McCann06}. Various consequences of
the opening of the gap were studied theoretically in
Refs.~\onlinecite{Blanter08, Stauber07, Impurities07, Min07,
0conductivity, Nilsson07, Benfatto08, Castro08, Guinea06, Sahu08,
exciton}. One of these consequences is that biased  bilayer
responds to disorder as a ``normal'' semiconductor, i.e.,
impurities give rise to the tails of the density of states which
extend into the gap from the bottom of conduction and from the top
of the valence band. Such in-gap localized states are especially
relevant to the experiment Ref.~\onlinecite{Ehop07}, where the
inelastic transport over these states has been observed. This
raises  a theoretical question about the shape of in-gap
fluctuation tails in bilayers and their dependence on the disorder
strength. The only paper on biased graphene bilayers with
impurities that we are aware of is Ref.~\onlinecite{Impurities07}.
This paper studies not the tails, but rather disorder-induced
smearing of the band-edges. Also, the numerical results for the
density of states in the gap region are presented in
Ref.~\onlinecite{Impurities07}  for particular values of impurity
concentration, $n_i$, so that the general dependence of the
magnitude of smearing on $n_i$, as well as on the interlayer bias,
$V$, was not established.

In the present paper we study analytically the density of
disorder-induced
localized in-gap states in bilayer graphene. The reason why
classical results \cite{Halperin66, Zittarz66, Thouless} for the
fluctuation tails do not directly apply to this situation, is a
peculiar structure of the bare energy spectrum shown in Fig.~1. In
particular, the minimum (maximum) of the electron (hole)
dispersion is located at finite momentum\cite{McCann06}, $p_0$.
Also, at energies of the order of the gap, the dispersion law is
not quadratic, but rather $\epsilon(p)\propto p^4$.  These two
features naturally define two regimes of disorder-induced
broadening of the density of states:

({\em i}) Weak disorder. The magnitude of smearing in this regime
is smaller than the depth of the minimum in Fig.~1. As a result,
the states responsible for the smearing have momenta close to
$p_0$. This fact, as was pointed out in
Ref.~\onlinecite{Impurities07}, facilitates localization of
electrons (holes) by weak impurities. Earlier this observation was
made in Refs. \onlinecite{RE88,we,Chaplik}.

({\em ii}) Strong disorder. The mexican-hat-structure in
Fig.~1 is completely smeared. In this regime, the in-gap
states are formed due to trapping of electrons (holes) with
{\em quartic} dispersion by certain disorder configurations.

\begin{figure}[t]
\centerline{\includegraphics[width=65mm,angle=0,clip]{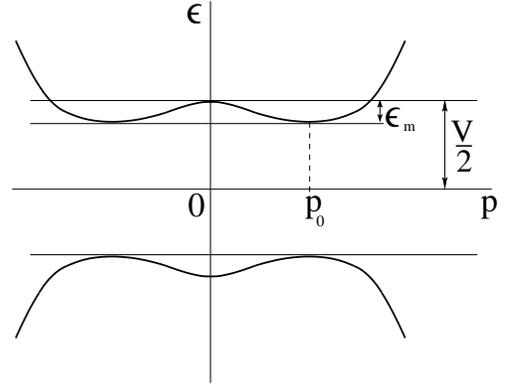}}
\caption{Energy spectrum of a biased bilayer graphene has a loop
of  minima
of depth $\epsilon_m$ Eq. (\ref{endeep}) at $\vert {\bf p}\vert=p_0$,
Eq. (\ref{p0}). At small bias, $V<t_{\perp}$, the gap is smaller than
the distance to the next subbands.}
\end{figure}

Below we demonstrate than in both regimes the magnitude of
smearing, $\epsilon_t$, is proportional to the combination
$n_i^{2/3}V^{1/3}$. Clearly, for the gap to be resolved, the
applied bias must exceed $\epsilon_t$. This suggests that the
threshold bias, for which $V>\epsilon_t$ is proportional to the
first power of $n_i$. We find the shape of the density of states
near the band-edges for two regimes by extending the instanton
approach of Refs.~\onlinecite{Halperin66, Zittarz66} to the
spectrum of a biased bilayer graphene. We restrict consideration
to the case of short-range impurity potential, $w({\bf r})$, so
that the correlator of the disorder potential is
\begin{eqnarray}
\label{correlator}
\langle U({\bf r})U({\bf r}^{\prime})\rangle= \gamma\delta({\bf
r}-{\bf r}^{\prime}),
\end{eqnarray}
with
\begin{eqnarray}
\gamma=n_i\left[\int d{\bf r}w({\bf r})\right]^2.
\end{eqnarray}
We also assume that the bias is smaller than the
interlayer tunneling constant \cite{Falko06,McCann06}, $t_{\perp}$.

\section{Bare density of states}

Due to interlayer hopping,
the spectrum of the
bilayer graphene becomes parabolic\cite{Falko06,McCann06},
$\epsilon(p)=\pm c^2p^2/t_{\perp}$, where $c$ is the Dirac velocity
in graphene. For
"small" momenta, $cp\ll t_{\perp}$, the gap opens upon applying the
bias, $V$, between the layers. For $V< t_{\perp}$, the low-energy
Hamiltonian of the bilayer graphene can be reduced
\cite{Blanter08,Manes07} to the $2\times2$ matrix

\begin{eqnarray}
\mathcal{H}=\left(\begin{array}{cc}
\frac V2\left(1-\frac{c^2p^2}{t_{\perp}^2}\right)&-\frac{c^2(p_x+ip_y)^2}{t_{\perp}}\\
\,\\ -\frac{c^2(p_x-ip_y)^2}{t_{\perp}}&-\frac
V2\left(1-\frac{c^2p^2}{t_{\perp}^2}\right)
\end{array}\right).
\end{eqnarray}
which yields the spectrum
\begin{eqnarray}
\epsilon^2(p)=
\frac{V^2}4\left(1-\frac{c^2p^2}{t^2_{\perp}}\right)^2
+\frac{c^4p^4}{t^2_{\perp}}.
\end{eqnarray}
It is seen that in addition to opening the gap, finite bias, $V$,
leads to negative effective mass at small momenta. Disorder
affects the energy domains $|\epsilon(p)\pm V/2|\ll V$, close to
the band edges. In these domains the spectrum can be further
simplified to
\begin{eqnarray}
\label{simpspec}
\epsilon^{\pm}(p)=\pm\left[ \frac{V}2+
\frac{c^4}{Vt_{\perp}^2}\left(p^2-p_0^2\right)^2\right],
\end{eqnarray}
where
\begin{eqnarray}
\label{p0}
p_0=\frac{V}{2c}
\end{eqnarray}
is the minimum position, Fig.~1. In Eq.~(\ref{simpspec}) we also
took into account that $V\ll t_\perp$. The eigenfunctions,
corresponding to the two branches, have the form
\begin{eqnarray}\label{ef}
\!\!\!\chi_{\bf p}^{+}({\bf r})\approx e^{i{\bf
pr}}\!\!\left(\!\!\begin{array}{c}
e^{2i\phi_{\bf p}}\\
\,\\-\frac {c^2p^2}{Vt_\perp}
\end{array}\!\!\right)\!;\;\;\;
\chi_{\bf p}^{-}({\bf r})\approx e^{i{\bf
pr}}\!\!\left(\!\!\begin{array}{c}
\frac{c^2p^2}{Vt_\perp}\\
\,\\ e^{-2i\phi_{\bf p}}
\end{array}\!\!\right)\!\!.
\end{eqnarray}
where $\phi_{\bf p}=\tan^{-1}(p_y/p_x)$ is the azimuthal angle of
the vector ${\bf p}$. The minimal energy for the branch
$\epsilon^+(p)$ is
\begin{eqnarray}
\Delta=\frac{V}{2}-\frac{V^3}{16t_{\perp}^2}.
\end{eqnarray}
Shifting the energy scale origin to $\Delta$, for the bare density
of states \cite{0conductivity}, $\nu_0(\epsilon)$, we will have
\begin{eqnarray}
\label{DOS}
\nu_0(\epsilon)=\left(\frac{4t^2_{\perp}}{Vc^2}\right)\tilde{\nu}
(\epsilon/\epsilon_m),
\end{eqnarray}
where
\begin{eqnarray}
\label{endeep} \epsilon_m= \frac{V^3}{16t_{\perp}^2},
\end{eqnarray}
is the depth of the minimum, Fig.~1, and the dimensionless
function $\tilde{\nu}(z)$ is defined as
\begin{eqnarray}
\label{dimless} \tilde{\nu}(z)= \left\{\begin{array}{l}
\frac1{\sqrt{z}},
\quad 0<z<1;\\
\,\\\frac1{2\sqrt{z}}, \quad z>1.
\end{array}\right.
\end{eqnarray}
Single-scale behavior of $\tilde{\nu}(z)$ ensures that the
magnitude of disorder-induced smearing of the band edges is
defined by a single parameter for both weak- and strong-disorder
regimes.

\section{Weak disorder}

In the vicinity of the minimum at $p=p_0$,
the dispersion law simplifies \cite{Impurities07, 0conductivity,
Guinea06} to
\begin{eqnarray}\label{mass}
\epsilon^{\pm}({\bf p})= \pm\frac{(|{\bf p}|-p_0)^2}{2m};\quad m=\frac
{t^2_{\perp}}{2Vc^2}.
\end{eqnarray}
This expansion applies in the domain, $(|{\bf p}|-p_0)\ll V/c$,
where $\epsilon({\bf p})-\epsilon(p_0)$ is smaller than the energy
distance $V^3/16t^2$ between the minimum and maximum in Fig.~1.
The eigenfunctions Eq.~(\ref{ef}) simplify to
\begin{eqnarray}\label{p0ef}
\!\!\!\chi_{\bf p}^{+}({\bf r})\approx e^{i{\bf
pr}}\!\!\left(\!\!\begin{array}{c}
e^{2i\phi_{\bf p}}\\
\,\\-\frac V{4t_\perp}
\end{array}\!\!\right)\!;\;\;\;
\chi_{\bf p}^{-}({\bf r})\approx e^{i{\bf
pr}}\!\!\left(\!\!\begin{array}{c}
\frac V{4t_\perp}\\
\,\\ e^{-2i\phi_{\bf p}}
\end{array}\!\!\right)\!\!.
\end{eqnarray}
One-dimensional character of the spectrum is reflected in the
$\epsilon^{-1/2}$ behavior of the density of states Eq.
(\ref{DOS}) near the band-edge. The magnitude, $\epsilon_t$, of
disorder-induced broadening can be estimated from
the following reasoning. The inverse scattering rate due to
the disorder, calculated from the golden rule, is given by
\begin{eqnarray}\label{tau}
\frac{1}{\tau(\epsilon)}=\frac{\gamma}{\pi}\,\nu_0(\epsilon)=
\frac{\gamma \,t_{\perp}V^{1/2}}{2\pi c^2\epsilon^{1/2}}.
\end{eqnarray}
Then the estimate for $\epsilon_t$ emerges upon equating
$1/\tau(\epsilon_t)$ to $\epsilon_t$, yielding
\begin{eqnarray}\label{tailenergy}
\epsilon_t=\left(\frac{V^{1/2}t_{\perp}}{c^2}\gamma\right)^{2/3}.
\end{eqnarray}
Note that the above consideration equally applies for weak
disorder, $\epsilon_t<\epsilon_m$, and strong disorder,
$\epsilon_t>\epsilon_m$. As follows from Eq.~(\ref{tailenergy}),
the weak-disorder regime realizes for
\begin{eqnarray}
\gamma \ll c^2\left(\frac V{t_{\perp}}\right)^4.
\end{eqnarray}

The fact that the fluctuation states in weak-disorder regime are
composed essentially from the free states with momenta, $p$, close
to $p_0$, allows to obtain an asymptotically exact solution for
the density of states in this regime. This was demonstrated in
Ref. \onlinecite{we}, where the spectrum Eq.~(\ref{mass}) emerged
as a result of the spin-orbit coupling. The difference between the
wave functions Eq.~(\ref{p0ef}) and spin-orbit wave functions
affects only numerical factors in the final result. Still, for
completeness, we will briefly sketch the calculation from Ref.
\onlinecite{we}, using our notations.

A drastic simplification coming from the condition
$\epsilon_t\ll\epsilon_m$ is that the interference between two
scattering amplitudes ${\bf p}\rightarrow{\bf p}^\prime$ and ${\bf
p}\rightarrow {\bf p}_1\rightarrow{\bf p}_2 \rightarrow{\bf
p}^\prime$ is suppressed. This is because the momenta ${\bf p}$,
${\bf p}_1$, ${\bf p}_2$, which are close to $p_0$ in {\em
absolute value}, are restricted in their mutual directions by the
condition $(|{\bf p}_1+ {\bf p}_2 -{\bf p}|-p_0)\sim
p_0(\epsilon_t/\epsilon_m)^{1/2}$. This condition limits the
angles between the momenta to $(\epsilon_t/\epsilon_m)^{1/2}$. As
a result, the non-random-phase-approximation (non-RPA) diagrams in
the self-energy, $\Sigma_{\bf p}(\epsilon)$, are parametrically,
in $(\epsilon_t/\epsilon_m)^{1/2}$, smaller than corresponding RPA
diagrams. In other words, the RPA becomes asymptotically exact in
weak-disorder regime.
\begin{figure}[t]
\centerline{\includegraphics[width=65mm,angle=0,clip]{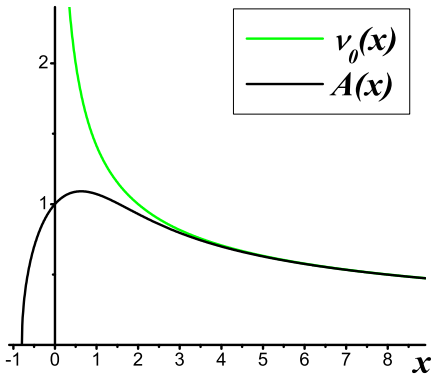}}
\caption{(Color online) Dimensionless density of states near the
band edge is plotted from Eq. (\ref{RPADOS}) versus dimensionless
energy $x=\epsilon/\epsilon_t$. Green line is the high-energy
asymptote, $\nu_0(x)= (2/x)^{1/2}$.}
\end{figure}

Within the RPA, the density of states is given by
\begin{equation}
\label{RPAdos} \nu (\epsilon) = \frac{1}{\pi }Im \sum _{\bm{p}}
\frac{|\chi^{(+)} _{\bf p}|^2}{\epsilon - \epsilon({\bf p})-
\Sigma_{{\bf p}}(\epsilon)},
\end{equation}
where the electron self-energy satisfies
\begin{equation}
\label{se} Im \Sigma_{\bf p}(E) = \gamma  Im \int \frac {d{\bf
p}_1}{(2 \pi)^2} \frac{|(\chi^{*(+)}_{\bf p} \chi^{(+)}_{{\bf
p}_1})|^2 }{ E - \epsilon({\bf p}_1) -\Sigma _{{\bf p}_1} (E) }.
\end{equation}
The fact that $Im \Sigma_{\bf p}$ does not depend on ${\bf p}$
allows to express the solution of Eq.~(\ref{se}) in the form
\begin{equation}
Im \Sigma(\epsilon) = \frac{\epsilon_t}{2^{5/3}} A\biggl
(\frac{2^{5/3} \epsilon}{\epsilon_t} \biggr ),
\end{equation}
where the energy $\epsilon$ is defined as
\begin{equation}
\epsilon = E-\left(\frac V2-\frac{V^3}{16t^2_\perp} + Re \Sigma
\right),
\end{equation}
and the dimensionless function $A(x)$ satisfies the algebraic
equation
\begin{equation}
A(x) = \sqrt{ \frac{x + \sqrt{A(x) ^{2}+x^{2} }} {A(x)
^{2}+x^{2}}}.
\end{equation}
The solution of this equation has the form
\begin{eqnarray}
\label{RPADOS} A(x)=\left(\left[\frac{3^{1/2}
2^{1/3}\left(3^{3/2}+\sqrt{27+4x^3}\right)^{1/3}}
{\left(3^{3/2}+\sqrt{27+4x^3}\right)^{2/3}-2^{2/3}x}\right]^2\!\!\!
-x^2\right)^{1/2}\!\!\! .\nonumber\\
\end{eqnarray}
The density of states per spin and per valley can be expressed via
$A(x)$ as
\begin{equation}
\nu(\epsilon)=\frac1{\pi\gamma}Im \Sigma(\epsilon) =
\frac{1}{2^{5/3}\pi}\frac{V^{1/2}t_\perp} {\epsilon_t^{1/2}c^2}
A\biggl (\frac{2^{5/3} \epsilon}{\epsilon_t} \biggr ).
\end{equation}
It is plotted in Fig.~2 together with the bare density of states.

Sharp low boundary of $\nu(\epsilon)$ in Fig.~2 is an artifact of
the RPA. In reality it is smeared within the energy interval
\begin{eqnarray}\label{newet}
\tilde{\epsilon}_t=\frac{\epsilon_t^{3/2}}{\epsilon_m^{1/2}}
=\frac{4\gamma t_\perp^2}{Vc^2}\ll \epsilon_t,
\end{eqnarray}
which is much smaller than $\epsilon_t$. This smearing comes from
non-RPA diagrams. Concerning the deep tail of $\nu(\epsilon)$, it
is close to a simple exponent, namely
\begin{eqnarray}\label{1dlike2d}
\nu(\epsilon)\propto
\exp\left[-\frac{4|\epsilon|}{\tilde{\epsilon}_t\ln(\epsilon_m/|\epsilon|)}\right].
\end{eqnarray}
The reason why this tail can be found analytically is again the
fact that the wave functions of the fluctuation states have two
spatial scales; they oscillate rapidly with period $p_0^{-1}$ and
decay at much larger distances as $\exp(-\sqrt{2m|\epsilon|}\,r)$.
The tail states are very similar to those found in Ref.
\onlinecite{we}. The key steps of derivation of
Eq.~(\ref{1dlike2d}) are outlined in the Appendix.

\section{Strong disorder}

We now turn to the case of a strong disorder when $\epsilon_t>\epsilon_m$.
The shape of the tail density of states in this case
can be established from the following
qualitative consideration. The probability density to
find a fluctuation $U({\bf r})$ is given by
\begin{eqnarray}
\label{prob0}\mathcal{P}\left\{U({\bf
r})\right\}=\exp\left[-\frac1{2\gamma}\int d{\bf r}\,U^2({\bf
r})\right].
\end{eqnarray}
In order to create a localized level with binding energy, $\epsilon$,
the magnitude of the fluctuation must exceed $\vert \epsilon\vert$,
while the size cannot be smaller than the de Broglie wave length,
$r_{\epsilon}$, of a free electron  with energy, $\epsilon$, i.e.,
\begin{eqnarray}
\label{rE}
r_{\epsilon}=\frac{1}{p_{\epsilon}}=
\left(\frac{c^4}{Vt_{\perp}^2|\epsilon|}\right)^{1/4},
\end{eqnarray}
where the last identity follows from the dispersion law
$\epsilon(p)=c^4p^4/Vt_{\perp}^2$. Now the integral
$\int d{\bf r}\,U^2({\bf r})$ can be estimated as
$\epsilon^2r_{\epsilon}^2$. Substituting this estimate
into Eq. (\ref{prob0}) and using Eq. (\ref{rE}), we get
\begin{eqnarray}
\label{tailDOS} \nu(\epsilon)\propto
\exp\left(-\Big|\frac{\epsilon}{\epsilon_t}\Big |^{3/2}\right).
\end{eqnarray}
The remaining task is to establish the numerical coefficient in
the exponent Eq. (\ref{tailDOS}) with the help of the instanton
approach\cite{Halperin66,Zittarz66,Thouless}. Within this
approach, one should solve the Schr\"{o}dinger equation with
potential, $U({\bf r})$, which yields an eigenvalue, $E$ (here we
measure energy  from the gap center). Then $U({\bf r})$ is
determined from the condition that $\int d{\bf r}\,U^2({\bf r})$
in the exponent of Eq. (\ref{prob0}) is minimal. This restriction
is conventionally incorporated by adding to $\int d{\bf
r}\,U^2({\bf r})$ the energy,
$E[\Psi]=\langle\Psi|(\mathcal{H}+U)|\Psi\rangle$ with Lagrange
multiplier, $\lambda$. Then minimization of
\begin{eqnarray}
\label{LagFnal} \lambda\langle\Psi|(\mathcal{H}+U)|\Psi\rangle -
\int d{\bf r}\,U^2({\bf r})
\end{eqnarray}
with respect to $U$ yields $U=-\frac\lambda2|\Psi|^2$.

At this point the following remark is in order.
Conventionally, upon substituting the found $U({\bf r})$
back into the Schr\"odinger equation,
the sign of $\lambda$ is chosen from the
condition that potential $U({\bf r})$ is attractive. However, in
our case of two symmetric bands, Fig.~1, the potential which is
attractive for electrons, is repulsive for holes, and vice versa.
It turns out that choosing $\lambda=2$ corresponds to $\nu(E)$
which falls off from $E=V/2$ {\it all the way} down to $E=-V/2$.
Correspondingly, choosing $\lambda=-2$ leads to the tail $\nu(-E)$
which grows from $E=-V/2$ towards the bottom of conduction band
$E=V/2$. Therefore, at $E=0$, we have two {\it different}
solutions, with $U({\bf r})$ and $-U({\bf r})$, for which the
"electron" and "hole" components of eigenfunction $\Psi$ are
related as $(\psi_e,\psi_h)\leftrightarrow
(-\psi_h^\ast,\psi_e^\ast)$. Thus, in view of exponential
character of $\nu(E)$, it is sufficient to set $\lambda=2$ and
consider only positive energies, $E>0$. The nonlinear instanton
equation reads
\begin{eqnarray}\label{instanton}
\frac
V2\psi_e-\frac{c^2}{t_{\perp}}\left(\partial_x+i\partial_y\right)^2\psi_h
-\psi_e\left(|\psi_e|^2+|\psi_h|^2\right)=E\psi_e;\nonumber\\
-\frac
V2\psi_h-\frac{c^2}{t_{\perp}}\left(\partial_x-i\partial_y\right)^2\psi_e
-\psi_h\left(|\psi_e|^2+|\psi_h|^2\right)=E\psi_h.\nonumber\\
\end{eqnarray}
Then, with exponential accuracy, we have $\nu(E)\propto
\mathcal{P}\left\{U({\bf r})\right\}$, where
\begin{eqnarray}
\label{probabil} \mathcal{P}\left\{U({\bf
r})\right\}=\exp\left[-\frac1{2\gamma}\int d{\bf r}\,U^2({\bf
r})\right]\\= \exp\left[-\frac1{2\gamma}\int d{\bf
r}\bigl(|\psi_e|^2+|\psi_h|^2\bigr)^2\right]\nonumber
\end{eqnarray}
is the probability of realization of $U({\bf r})$.

Consider first the energies close to the bottom of conduction
band, $-\epsilon=\left(\frac V2-E\right)\ll V$. In this limit, the
second equation in the system Eq. (\ref{instanton}) can be
simplified as
\begin{eqnarray}
\label{simpl}
\psi_h=-\frac{c^2}{Vt_{\perp}}\left(\partial_x-i\partial_y\right)^2\psi_e.
\end{eqnarray}
Substituting Eq.~(\ref{simpl}) into the first equation Eq.
(\ref{instanton}), and performing rescaling
\begin{eqnarray}
\label{resc} r= c(Vt_{\perp}^2|\epsilon|)^{-1/4}\rho,\quad
\psi_e=|\epsilon/\lambda|^{1/2}f(\rho),
\end{eqnarray}
we arrive to the following dimensionless instanton equation
\begin{eqnarray}
\label{die}\Delta_\rho^2f(\rho)+f(\rho)-f(\rho)^3=0,
\end{eqnarray}
while the expression for $\nu(\epsilon)$ takes the form
\begin{eqnarray}
\label{edgeden}\nu(\epsilon)\propto\exp\left[-\frac
{I_4}2\bigg|\frac\epsilon{\epsilon_t}\bigg|^{3/2}\right],
\end{eqnarray}
where $\epsilon_t$ is defined by Eq. (\ref{tailenergy}) and
$I_4=\int d{\bm{\rho}} f^4(\rho)$.
\begin{figure}[t]
\centerline{\includegraphics[width=65mm,angle=0,clip]{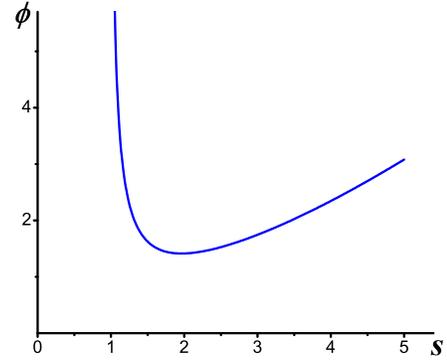}}
\caption{(Color online) The function $\phi(s)$ is plotted from
Eq.~(\ref{mf}).}
\end{figure}

Eq.~(\ref{instanton}) has solutions for arbitrary angular
momentum, $M$. However, leading contribution to the density of
states comes from azimuthally symmetric solution, $f(\rho)$. Then
the hole component Eq. (\ref{simpl}) of the wave function
corresponds to the momentum ${\cal M}=2$, namely
\begin{eqnarray}
\label{hcomp}
\psi_h=\frac{\epsilon}{\sqrt{V\lambda}}e^{-2i\phi}\left(\partial_\rho^2
-\frac1\rho\,\partial_\rho\right)f(\rho).
\end{eqnarray}
A peculiar feature of the non-linear equation Eq.~(\ref{die}) is
that it contains $\Delta_\rho^2$ instead of a usual Laplace
operator, $\Delta_\rho$. This is a direct consequence of the
dispersion law $\epsilon(p)\propto p^4$. As a result, the average
"kinetic" energy, $J_2=\int d{\bm{\rho}}[\Delta_\rho f(\rho)]^2$,
"potential" energy $I_4$ and the integral $I_2=\int
d{\bm{\rho}}[f(\rho)]^2$ are related as $1\!:\!4\!:\!3$, unlike
the relation $1\!:\!2\!:\!3$ for conventional polaron
\cite{Pekar}. We solve Eq.~(\ref{die}) by employing the
variational approach on the class of trial functions
$f(\rho)=C\exp[-(\rho/\rho_0)^s]$. Minimization of the
corresponding functional
\begin{eqnarray}
\label{va1}\mbox{\Large $\Phi$}\left\{f \right\}=\int
d{\bm{\rho}}\left[(\Delta_\rho f)^2 + f^2-\frac12f^4\right]
\end{eqnarray}
with respect to $C$ and $\rho_0$ can be easily performed
analytically. The resulting $s$-dependence of
$\Phi\left\{f\right\}$ has the form
\begin{figure}[t]
\centerline{\includegraphics[width=65mm,angle=0,clip]{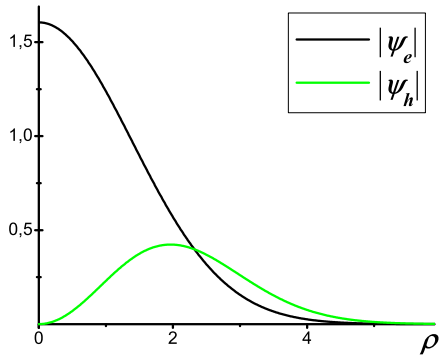}}
\caption{(Color online) Electron and hole components of the
wave function, obtained from variational solution of Eq.~(\ref{instanton})
for $E=0$, are plotted versus dimensionless distance, $\rho$, defined by
Eq. (\ref{resc}).}
\end{figure}
\begin{eqnarray}
\label{mf}\phi(s)=\left[\pi\frac{s(s-1)+2}{2^{3-4/s}}
\frac{(1-2/s)}{\sin\pi(1-2/s)}\right]^{1/2}.
\end{eqnarray}
This combination has a well-pronounced minimum at $s\approx 1.963$
see Fig.~3.
On the other hand, the "hole" wave function Eq. (\ref{hcomp}) is
non-singular at $\rho\rightarrow 0$ if $s\geq 2$. Thus it is
reasonable to adopt $s=2$, yielding $C=1.63$ and $\rho_0=2.21$.
This allows to specify the numerical factor in the exponent of
density of states in Eq.~(\ref{edgeden}), namely $I_4= 27.36$.

The result Eq.~(\ref{edgeden}) applies for $\epsilon_t\ll
\epsilon\ll V$. A relevant question for inelastic transport is the
density of states at the gap center. For qualitative estimate it
is sufficient to substitute $\epsilon=-V/2$ into the
Eq.~(\ref{edgeden}), which recovers Eq.~(\ref{rho0}). To establish
the numerical coefficient in the exponent more accurately, we
found variational solution of the system Eq.~(\ref{instanton}) for
$E=0$. It turns out that the coefficients $C$ and $\rho_0$ in
$f(\rho)$ assume the values $1.6$ and $1.96$ respectively, and the
numerical coefficient in Eq.~(\ref{edgeden}) is $I_4=23.22$. We
conclude that the expression Eq.~(\ref{edgeden}) for the tail of
the density of states is essentially valid down to the gap
center.

\section{Concluding remarks}

\noindent{\bf 1.} {\em Prefactor.}
Eq. (\ref{edgeden}) describes the tail of the
density of states with exponential accuracy. To restore the
dimensionality, it is natural to multiply Eq. (\ref{edgeden})
by $\nu_0(\epsilon_t)$, since the smearing of the band-edge is
$\sim\epsilon_t$. However, since Eq. (\ref{edgeden}) describes the
deep tail, the prefactor contains an additional power of
a dimensionless ratio  $(|\epsilon|/\epsilon_t)$. To establish
this power, one has to follow the procedure of calculating
the prefactor in the functional integral \cite{Parisi80, ER}.
Within this procedure, the origin of a prefactor is the
fact that the center of the instanton can be shifted
in the plane along both axes. These shifts correspond
to the so-called zero modes. Each zero mode contributes a
factor $(|\epsilon|/\epsilon_t)^{1/4}$. The power $1/4$
reflects the size of the instanton fluctuation,
$r_{\epsilon}\propto |\epsilon|^{-1/4}$, see Eq.~(\ref{rE}).
Overall, within a numerical coefficient, the final expression
for the density of states in the tail reads
\begin{eqnarray}
\label{finDOS}
\nu(\epsilon)=\frac{V^{1/2}t_\perp}{c^2\epsilon_t}
|\epsilon|^{1/2}
\exp\left[-11.6\bigg|\frac\epsilon{\epsilon_t}\bigg|^{3/2}\right].
\end{eqnarray}
Note that, unlike the case of parabolic spectrum\cite{Parisi80},
for $\epsilon(p)\propto p^4$, the prefactor does not diverge.
This is because the second-order shift of the band-edge
$\propto \gamma \int d\epsilon \nu_0(\epsilon)/\epsilon$
converges at large $\epsilon$ for
$\nu_0(\epsilon) \propto \epsilon^{-1/2}$.

\noindent{\bf 2.} {\em Relation to the scattering time.}
Short-range disorder is characterized by a single parameter,
$\gamma$. For comparison with experiment, this parameter can be
related to electron scattering time, $\tau_{\scriptscriptstyle F}$,
in the case
when the gate voltage places the Fermi level, $E_F$,
well above the smearing, $\epsilon_t$, of the band-edge.
Expressing $\gamma$ from Eq. (\ref{tau}), and substituting
into Eq. (\ref{tailenergy}), we obtain
\begin{eqnarray}
\label{tailviatau}
\epsilon_t=\left(\frac{4\pi^2E_{\scriptscriptstyle F}}
{\tau_{\scriptscriptstyle F}^2}\right)^{1/3}
=\left(\frac{4\pi^4n_e^2c^4}{Vt_\perp^2\tau_{\scriptscriptstyle F}^2}
\right)^{1/3}.
\end{eqnarray}
In the second identity we had expressed $E_{\scriptscriptstyle F}$ via
electron density, $n_e$.

\noindent{\bf 3.} {\em The role of intervalley scattering.}
Our assumption that disorder is short-ranged requires that
the radius of the impurity potential, $w(r)$, is smaller
than the wavelength of electron with energy $\sim V$, which
is $\sim c/\left(Vt_{\perp}\right)^{1/2}$. This length is
much larger than the lattice constant, and we neglected
the intervalley scattering. If the radius of $w(r)$ is
comparable to the lattice constant, the intervalley scattering
becomes as efficient as intravalley scattering. This would not
only lift the valley degeneracy of the fluctuation states
but also result in their azimuthal asymmetry, much like in
the case of degenerate valence band
considered in Ref. \onlinecite{kusmartsev}.
The consequence of this asymmetry is the change of
the numerical factor in the exponent of
Eq.~(\ref{edgeden}).

\noindent{\bf 4.} {\em Dependence on impurity concentration.}
Our main result Eq.~(\ref{finDOS}) applies in the limit
of strong disorder when $\gamma > c^2(V/t_{\perp})^4$.
On the other hand, we assumed that the gap is not washed
out completely by the disorder. Then the upper limit on $\gamma$
can be found by setting $\epsilon=V/2$ in Eq.~(\ref{finDOS})
equating the exponent to $1$.
Finally, it is convenient to present the domain of validity of
Eq.~(\ref{finDOS}) as
\begin{eqnarray}
\label{validity}
\left(\frac V{t_{\perp}}\right)^4<\frac \gamma{c^2}<4.1\frac V{t_{\perp}}.
\end{eqnarray}
It follows from the second condition that
the minimal $V=V_c$, at which the gap effectively opens,
is proportional to the impurity concentration, as it
was stated in the Introduction.
We can also rewrite the above
condition in terms of dimensionless conductance when the Fermi
level is in the conduction band
\begin{eqnarray}
E_{\scriptscriptstyle F}\tau_{\scriptscriptstyle F}>
1.5\left(\frac{E_{\scriptscriptstyle F}}{V}\right)^{3/2}.
\end{eqnarray}
Finally, we address the case of low impurity
concentration when the first condition Eq.~(\ref{validity})
is violated.

An isolated impurity with potential, $w(r)$, creates
a localized state with binding energy \cite{RE88,we,Chaplik}
\begin{eqnarray}
\label{bindenergy}
\epsilon_b=2\pi^2mp_0^2\left(\int d{\bf r} w(r)J_0^2(p_0r)\right)^2.
\end{eqnarray}
For the short-range potential, $w(r)$, the Bessel function in the
integrand can be set to $1$. Substituting $p_0$ and $m$ from Eqs.
(\ref{p0}) and (\ref{mass}), we obtain
\begin{eqnarray}
\label{beV}
\epsilon_b=\frac{\pi^2Vt_\perp^2}{4c^2}\left(\int d{\bf r} w(r)\right)^2.
\end{eqnarray}
We see that, for bilayer graphene,
the binding  energy is proportional to the gap \cite{Impurities07}.
The wave function of this localized state not only
falls off exponentially with distance, $r$, from the
impurity, but also oscillates
as $J_0(p_0r)$. It is clear that when the average
distance between the impurities $\sim n_i^{-1/2}$
becomes smaller than $p_0^{-1}$ the impurity band
merges with the conduction (valence) band.
Remarkably, the criterion
\begin{eqnarray}
\label{CONDITION}
n_i>p_0^2
\end{eqnarray}
also follows from a very different reasoning.
The expression Eq.~(\ref{newet}) for the tail
energy, $\tilde{\epsilon}_t$, in the weak-disorder
regime can be rewritten as
\begin{eqnarray}
\label{rewritten}
\tilde{\epsilon}_t=\frac{4n_it_\perp^2}{Vc^2}
\left(\int d{\bf r} w(r)\right)^2.
\end{eqnarray}
Comparing Eq.~(\ref{rewritten}) to Eq.~(\ref{bindenergy}),
we find that the ratio $\tilde{\epsilon}_t/\epsilon_b$
is $\sim n_i/p_0^2$. Thus the condition Eq. (\ref{CONDITION})
that neighboring localized states overlap ensures that
the impurity band transforms into the tail of the
conduction (valence) band. Note also, that in treating
the weak-disorder regime we assumed that the correlator
of the disorder potential is given by Eq. (\ref{correlator}).
By making this assumption we already implied that disorder is not
due to individual impurities but rather due to fluctuations
of impurity concentration, i.e., that the levels
Eq.~(\ref{bindenergy})
are not formed at $n_i$ satisfying the condition
Eq.~(\ref{CONDITION}). In conclusion, we rewrite
for completeness the
condition of strong disorder in terms of impurity
concentration and binding energy of an individual
impurity
\begin{eqnarray}
\label{finalcondition}
\frac{V^3}{\epsilon_bt_\perp^2}<\frac{n_i}{\pi^2p_0^2}
<4.1\frac{t_\perp}{\epsilon_b}.
\end{eqnarray}
We see that the smaller is the gap,  the broader is
the interval Eq.~(\ref{finalcondition}).

\section{Acknowledgement}
We are grateful to A. K. Savchenko for initiating this work. The
work was supported by the Petroleum Research Fund under Grant No.
43966-AC10.

\appendix

\section{}
Similar to Eq.~(\ref{probabil}), $\nu (\epsilon)$ in the tail is
given by
\begin{equation}
\label{roe} \nu (\epsilon) \propto \exp \biggl (- \frac{1}{2
\gamma}\int d {\bf r}|\varphi ({\bf r})|^4 \biggr ) ,
\end{equation}
where the two-component function $\varphi ({\bf r})$ satisfies the
equation
\begin{equation}
\label{instantoneq} \hat H \varphi ({\bf r}) -| \varphi ({\bf
r})|^2 \varphi ({\bf r}) =\epsilon \, \varphi ({\bf r}),
\end{equation}
with $\hat H$ being the free Hamiltonian with the spectrum
Eq.~(\ref{mass}) and the eigenfunctions Eq.~(\ref{p0ef}).
Searching the solution in the form
\begin{equation}\label{FT}
\varphi ({\bf r}) =  \int d {\bf p} \!\   B({\bf p}) \chi^{+}_{\bf
p}({\bf r}),
\end{equation}
we arrive to the following integral equation for $B({\bf p})$:
\begin{eqnarray}
\label{ak1} && B({\bf p})\bigl[\epsilon^+({\bf p})-\epsilon\bigr]=
\frac{1}{(2 \pi )^{2}} \int\!\! d{\bf r}
 \left(\int \prod_{i=1,2,3}d {\bf p}_iB({\bf p}_i)\right) \nonumber \\
&&~~~~~~~~~~~~~~~~~~~~~\times \Bigl(\chi^{+*}_{{\bf p}}
\chi^{+}_{{\bf p}_1}\Bigr) \Bigl(\chi^{+*}_{{\bf p}_2}
\chi^{+}_{{\bf p}_3}\Bigr).
\end{eqnarray}
Assuming that $B({\bf p})$ depends only on the absolute value of
${\bf p}$, we easily perform the angular integration in
(\ref{ak1}). Using explicit forms of the wave function scalar
products, we obtain
\begin{eqnarray}
\label{ak2} && B({\bf p})\bigl[\epsilon^+({\bf p})-\epsilon\bigr]=
 \pi ^{2}\!\! \int\!\! dr\, r
 \left(\int \prod_{i=1,2,3}d {\bf p}_iB({\bf p}_i)\right) \nonumber \\
&&~~~~~~\times J_{2}(pr)J_{2}(p_1r)J_{2}(p_2r)J_{2}(p_3 r),
\end{eqnarray}
where $J_2 (x)$ is the Bessel function of the second order. The
product of $J_2 (p_ir)$ manifests the difference of Eq.
(\ref{ak2}) from the corresponding equation in Ref.
\onlinecite{we}.

The principal step in solving Eq. (\ref{ak2}) is setting all
momenta in the right hand side equal to $p_0$. Then the integral
over $r$ yields $\frac12\ln(\epsilon_m/|\epsilon|)$ so that Eq.
(\ref{ak2}) reduces to
\begin{equation}
\label{ak3}  B({\bf
p})\!\left[\frac{(p-p_0)^2}{2m}-\epsilon\right]\!=\! 2p_0
\ln\left(\frac{\epsilon_m
}{|\epsilon|}\right)\!\left[\int_{0}^{\infty}\!\!d
p^{\prime}B(p^{\prime}) \right]^3\!\!\! .
\end{equation}
This equation has an obvious solution of the form
\begin{equation}
\label{solution} B(p)= \frac{\beta}{(p-p_0)^2/2m +|\epsilon |} \ .
\end{equation}
Substituting Eq.~(\ref{solution}) into Eqs. (\ref{ak3}) and
(\ref{FT}), we find for constant $\beta$ the value
\begin{equation}
\label{const} \beta=\frac{1}{2^{1/2} \pi^{3/2}} p^{-1/2}_0  \left
(\frac{2m}{|\epsilon|}\right )^{-3/4} \ln ^{-1/2}(\epsilon_m /
|\epsilon|),
\end{equation}
and for $\varphi (r)$ the form
\begin{equation}
\label{fi} \varphi (r)= 2 \pi^2\beta p_0\left (\frac{m}{
|\epsilon|}\right)^{1/2} \left( \begin{array}{c} J_2(p_0r)\\ \,\\
0
\end{array}\right),
\end{equation}
which is valid for $r \lesssim (2m|\epsilon|)^{-1/2}$.  Finally,
Eq.~(\ref{1dlike2d}) emerges upon substituting Eq.~(\ref{fi}) into
Eq.~(\ref{roe}).

\end{document}